\documentstyle[12pt]{article}

\newcommand {\nn}    {\nonumber}
\newcommand {\vs}[1]  { \vspace*{#1 cm} }

\newcounter{eq}
\newcounter{sc}

\newcommand {\MPL}  {Mod.Phys.Lett.}
\newcommand {\NP}   {Nucl.Phys.}
\newcommand {\PL}   {Phys.Lett.}
\newcommand {\PR}   {Phys.Rev.}

\newcommand {\JHEP}  {J.High Energy Phys.}

\def\overleftrightarrow#1{\vbox{\ialign{##\crcr
 $\leftrightarrow$\crcr\noalign{\kern-1pt\nointerlineskip}
 $\hfil\displaystyle{#1}\hfil$\crcr}}}

\newlength{\minitwocolumn}
\begin{document}

\begin{flushright}
EDO-EP-24\\
October, 1998\\
\end{flushright}
\vspace{30pt}

\pagestyle{empty}
\baselineskip15pt

\begin{center}
{\large\bf $SL(2,Z)$ S-duality of Super D-string \\
in \\
Type IIB Supergravity Background
 \vskip 1mm
}

\vspace{20mm}

Ichiro Oda
          \footnote{
          E-mail address:\ ioda@edogawa-u.ac.jp
                  }
\\
\vspace{10mm}
          Edogawa University,
          474 Komaki, Nagareyama City, Chiba 270-0198, JAPAN \\

\end{center}


\vspace{15mm}
\begin{abstract}
It is shown in a quantum-mechanically exact manner
that a supersymmetric and $\kappa$-symmetric D-string 
action in a general type IIB supergravity background is transformed 
to a form of the type IIB Green-Schwarz superstring action with the 
$SL(2,Z)$ covariant tension through an S-duality transformation.
This result precisely proves a conjecture mentioned previously 
that the $SL(2,Z)$ S-duality of a super D-string action 
in a flat background is also valid even in a curved IIB background 
geometry. We point further out the validity of the more generalized 
conjecture that various duality relations of super D-brane and M-brane
actions originally found in a flat background also hold true in general
ten dimensional type II supergravity and eleven dimensional 
supergravity background geometries by applying the present formalism 
to those cases.
\vspace{15mm}

\end{abstract}

\newpage
\pagestyle{plain}
\pagenumbering{arabic}


\rm
\section{Introduction}

Symmetries have always played a central role in theoretical
physics. For instance, gauge symmetries are used not only
to determine dynamics, the gauge interactions, but also
to prove unitarity and renormalizability of a theory.
Thus the importance of a knowledge of symmetries in
modern theoretical physics is recognized as one of key
ingredients in establishing a theory.

However, in superstring theory and M-theory, symmetries
have remained elusive so that many vital issues have had to
be left to await the advent of an understanding of symmetries.
Such an understanding has begun to emerge recently owing to
developments of some non-perturbative techniques although
we are still quite a way (it seems) from addressing what 
ultimate symmetries in superstring theory and M-theory are.

In this respect, over recent years a large amount of evidence
has accumulated in support of various duality relations of
superstring theory and M-theory. Following the evidence,
these theories have a huge group of global discrete symmetries.
A particularly interesting subgroup is the $SL(2,Z)$ S-duality
\cite{Schwarz} of type IIB superstring theory \cite{GS}
which exchanges an infinite family of solitons and bound states. 
It was already known that the IIB supergravity 
\cite{Schwarz2, Howe}, which is the low energy effective theory 
of the type IIB superstring theory, possesses a discrete
global symmetry, the $SL(2,R)$ symmetry, but this symmetry was 
regarded as an artifact of the low energy approximation to
underlying renormalizable theory and not taken seriously in 
those days.
But nowadays the situation has radically changed since the 
discovery of the still rather mysterious more fundamental theory 
where $SL(2,Z)$ subgroup of $SL(2,R)$ is expected to be an 
exact symmetry \cite{Hull}. Hence it is an important and
urgent issue to obtain a fuller understanding of the $SL(2,Z)$ 
symmetry when one tries to clarify aspects of symmetries
behind the more fundamental underlying theory.

In this paper, we wish to consider a derivation of the $SL(2,Z)$
S-duality \cite{Schwarz} of type IIB superstring theory in a 
general type IIB supergravity background \cite{Cederwall1, 
Bergshoeff}. 
The conjecture that the $SL(2,Z)$ S-duality \cite{Schwarz} 
of type IIB superstring theory and other dualities of D-brane 
and M-brane actions, which were found originally in a flat background 
may be true even in a general curved on-shell background, was 
already stated in the final section of the papers 
\cite{Aganagic1, Cederwall2}, but only recently was shown in the
case of super D-string and super D3-brane actions on a specific 
$AdS_5 \times S^5$ background where it was strongly suggested 
that this conjecture would be valid even in a general on-shell 
supergravity background \cite{Oda1, Oda2}
(See also an interesting work \cite{Kimura} where the gauge-fixing
of $\kappa$-symmetry was not used).

The main purpose of this paper is to report that this 
conjecture is indeed the case. Because of the character as a
short article, we confine ourselves to only the case of a
super D-string action in a general type IIB supergravity 
background and present essential ideas and techniques to
some extent, 
but it is quite straightforward to extend the ideas and techniques
presented in this paper to the broader situations, namely a 
proof of various duality properties of super D-brane actions 
in ten dimensional type IIA and IIB supergravity background, 
and of M2 and M5-brane actions in eleven dimensional supergravity 
background such as $SL(2,Z)$ self-duality of D3-brane, 
D2 vs. M2-brane duality and D4 vs. M5-brane duality. 
These problems will be reported in a longer paper in detail
\cite{Oda3}. 

In this paper, as method of derivation of the $SL(2,Z)$ S-duality
in a super D-string action, we rely on path integral of the 
first-order Hamiltonian form, which was used in \cite{de Alwis} 
for a bosonic string and in \cite{Oda4} for a superstring in a 
flat Minkowskian background, and also in \cite{Oda1} for a 
superstring on $AdS_5 \times S^5$.

The contents of this article are organized as follows. 
In Section 2 we shall review a super D-string action in a
general IIB supergravity background \cite{Cederwall1, Bergshoeff}.
In Section 3 it will be shown that the super D-string action on 
this background is transformed to the type IIB Green-Schwarz 
superstring action with the $SL(2,Z)$ covariant tension in the 
same background through an S-duality transformation.
The final section will be devoted to discussions.

\section{ Super D-string action in a general IIB background}

We start by reviewing a super D-string action in a general 
IIB supergravity background \cite{Cederwall1, Bergshoeff}.
It is well known nowadays that super D-brane actions consist
of two terms, those are, the Dirac-Born-Infeld action and
the Wess-Zumino action. The former includes the NS-NS two-form and 
dilaton in addition to world-volume metric while the latter
action contains the coupling of the D-brane to the R-R fields.
In particular, a super D-string action in a general IIB on-shell 
supergravity background is given by
\begin{eqnarray}
S = S_{DBI} + S_{WZ},
\label{2.1}
\end{eqnarray}
with
\begin{eqnarray}
&{}& S_{DBI} = -  \int_{M_2 = \partial M_3} d^2 \sigma 
\sqrt{- \det ( G_{ij} + {\cal F}_{ij} )}, \nn\\
&{}& S_{WZ} =  \int_{M_2 = \partial M_3} C_2 = \int_{M_3} 
\Omega_3, \nn\\
\label{2.2}
\end{eqnarray}
where we have defined as
\begin{eqnarray}
{\cal F} &=& F - b_2, \nn\\
F &=& dA, \nn\\
\Omega_3 &=& dC_2.
\label{2.3}
\end{eqnarray}
Moreover, following the paper \cite{Cederwall1}, we define the NS-NS
3-form superfield $H_{(3)}$ of the NS-NS 2-form $b_2$ and 
the R-R $n$-form superfield $R$ as 
\begin{eqnarray}
H_{3} &=& db_2, \nn\\
R &=& e^{b_2} \wedge d( e^{-b_2} \wedge C )
= \displaystyle \bigoplus_{n=1}^{10} R_{(n)}, \nn\\
C &=& \displaystyle \bigoplus_{n=0}^{9} C_{(n)},
\label{2.4}
\end{eqnarray}
but in the case of string, it is easy to show that the R-R 3-form 
field strength superfield $R_{(3)}$, which is relevant below, 
coincides with the Wess-Zumino form $\Omega_3$. 
Of course, from the definitions (\ref{2.4}) these field strengths 
obey the following Bianchi identities
\begin{eqnarray}
dH_{3} &=& 0, \nn\\
e^{b_2} \wedge d( e^{-b_2} \wedge R )
= dR - R \wedge H_{3} &=& 0.
\label{2.5}
\end{eqnarray}

In order to reduce the enormous unconstrained field content
included in the superfields to the field content of the on-shell 
type IIB supergravity theory, one has to impose the constraints 
on the field strengths by hand, which make various Bianchi identities
to coincide with the equations of motion of supergravity
\cite{Howe}.
In this section, for the sake of clarity, we shall limit ourselves 
to be the case of the vanishing dilaton superfield. Later we will 
consider the constant dilaton background when we derive the $SL(2,Z)$
covariant tension. 
The case of a general dilaton superfield will be discussed at the
end of next section.
Under the assumption of the vanishing
(or constant) dilaton, the nontrivial constraints imposed on the field 
strength components \cite{Cederwall2} reduce to simple forms
\begin{eqnarray}
T_{\alpha\beta}^c &=& 2i \gamma_{\alpha\beta}^c, \nn\\
H_{a \alpha\beta} &=& - 2i ({\cal K} \gamma_a)_{\alpha\beta}, \nn\\
\Omega_{a \alpha\beta} &=&  2i ({\cal I} \gamma_a)_{\alpha\beta},
\label{2.6}
\end{eqnarray}
where $T_{\alpha\beta}^c$ indicates a component of the torsion
superfield $T_{ABC}$. And ${\cal E}$, ${\cal I}$, and ${\cal K}$  
describing the $SO(2)$ matrices are defined in terms of the
conventional Pauli matrices
\begin{eqnarray}
{\cal E} = i \sigma_2 = \pmatrix{
0  & 1 \cr -1 & 0 \cr }, \ {\cal I} = \sigma_1 = \pmatrix{
0  & 1 \cr 1 & 0 \cr },  \ {\cal K} = \sigma_3 = \pmatrix{
1  & 0 \cr 0 & -1 \cr }.
\label{2.7}
\end{eqnarray}

{}From these constraints, it is easy to
obtain the NS-NS 3-form superfield $H_{3}$ and the 3-form
Wess-Zumino term $\Omega_3$ whose concrete expressions are
given by
\begin{eqnarray}
H_{3} &=& i \bar{E} \wedge \hat{E} \wedge {\cal K} E, \nn\\
\Omega_3 &=& - i \bar{E} \wedge \hat{E} \wedge {\cal I} E,
\label{2.8}
\end{eqnarray}
where $E$ implies an appropriate component of the vielbein one-form 
$E^A = dZ^A E_M^A$. Writing down in an explicit manner, 
$\bar{E}$, $\hat{E}$ and $E$ in  
(\ref{2.8}) indicate the Dirac adjoint of $E^{I\alpha}$, $E^a \gamma_a$
and $E^{I\alpha}$, respectively. Here we have defined the curved
super-index as $M = (m, \mu)$ and the flat super-index $A = 
(a, \alpha)$, and $I$ as the $N=2$ index taking 1 and 2.

For later convenience, let us recapitulate the results obtained
in this section. The $\kappa$-symmetric and world-sheet 
reparametrization invariant super D-string action in a IIB 
on-shell supergravity background is given by
\begin{eqnarray}
S &=& S_{DBI} + S_{WZ}, \nn\\
S_{DBI} &=& - \int_{M_2} d^2 \sigma 
\sqrt{- \det ( G_{ij} + {\cal F}_{ij} )}, \nn\\
S_{WZ} &=& \int_{M_2 = \partial M_3} C_2 = \int_{M_3} 
\Omega_3, \nn\\
{\cal F} &=& F - b_2, \nn\\
H_{3} &=& db_2 = i \bar{E} \wedge \hat{E} \wedge {\cal K} E, \nn\\
\Omega_3 &=& dC_2 = - i \bar{E} \wedge \hat{E} \wedge {\cal I} E.
\label{2.9}
\end{eqnarray}
At this stage, it is interesting to notice that these equations
are quite similar to those of a super D-string action on 
$AdS_5 \times S^5$ \cite{Oda1} whose observation actually
made it possible to achieve the present study.

\section{ SL(2,Z) S-duality}

Now let us turn our attention to a proof of $SL(2,Z)$ S-duality of 
a super D-string action in a general ten dimensional IIB supergravity
background. As mentioned just above, since the action is written to be
a form similar to that on $AdS_5 \times S^5$ \cite{Oda1}, we can
follow a similar path of derivation, but we will expose the procedure
to some extent for completeness. The main focus is paid on the point
of how one performs the $SO(2)$ spinor rotation in moving from
the super D-string action to the Green-Schwarz action 
through the othogonalization of the $SO(2)$ matrix.

Now we are ready to show how the super D-string action 
(\ref{2.9}) becomes a fundamental Green-Schwarz superstring action  
with the $SL(2, Z)$ covariant tension by using the path integral
of the first-order Hamiltonian form \cite{de Alwis, Oda4, Oda1}.

According to the Hamiltonian formalism, let us start by introducing
the canonical conjugate momenta $\pi^i$ corresponding to the 
gauge field $A_i$ defined as
\begin{eqnarray}
\pi^i \equiv \frac{\partial S}
{\partial \dot{A}_i} =  \frac{\partial S_{DBI}}
{\partial \dot{A}_i},
\label{3.1}
\end{eqnarray}
where we used the fact that the Wess-Zumino term is independent
of the gauge potential, which holds only in the case of string
theory. Then the canonical conjugate momenta $\pi^i$ are calculated to
be
\begin{eqnarray}
\pi^0 = 0, \ \pi^1 = \frac{{\cal F}_{01}}
{\sqrt{-\det ( G_{ij} + {\cal F}_{ij} )}},
\label{3.2}
\end{eqnarray}
where the former equation just shows the existence of the $U(1)$
gauge invariance.  From these equations we will see that the 
Hamiltonian density takes the form 
\begin{eqnarray}
{\cal H} = \sqrt{1 + (\pi^1)^2} \sqrt{- \det G_{ij}} 
- A_0 \partial_1 \pi^1 + \partial_1 (A_0 \pi^1)
+ \pi^1 b_{01} - C_{01},
\label{3.3}
\end{eqnarray}
 
Now the partition function is defined by the first-order 
Hamiltonian form with respect to only the gauge field as follows:
\begin{eqnarray}
Z &=& \frac{1}{\int {\cal D}\pi^0} \int {\cal D}\pi^0
{\cal D}\pi^1 {\cal D}A_0 {\cal D}A_1 
\exp{ i \int d^2 \sigma ( \pi^1 \partial_0 A_1 - {\cal H} ) } \nn\\
&=& \int {\cal D}\pi^1 {\cal D}A_0 {\cal D}A_1 
\exp{ i \int d^2 \sigma} \nn\\
& & {} \times \left[ - A_1 \partial_0 \pi^1 + A_0 \partial_1
\pi^1 -  \sqrt{1 + (\pi^1)^2} \sqrt{- \det G_{ij}} 
- \pi^1 b_{01} + C_{01} - \partial_1 ( A_0 \pi^1 ) \right].
\label{3.4}
\end{eqnarray}

Provided that we take the boundary conditions for $A_0$ such that 
the last surface term in the exponential identically vanishes, 
then we can carry out the integrations over $A_i$ explicitly, 
which gives rise to $\delta$ functions
\begin{eqnarray}
Z &=& \int {\cal D}\pi^1 \delta(\partial_0 \pi^1) 
\delta(\partial_1 \pi^1) \exp{ i \int d^2 \sigma 
\left[ -\sqrt{1 + (\pi^1)^2} \sqrt{- \det G_{ij}} 
+ C_{01} - \pi^1 b_{01} \right] }.
\label{3.5}
\end{eqnarray}
The existence of the $\delta$ functions reduces the integral 
over $\pi^1$ to the one over only its zero-modes. If we require
that one space component is compactified on a circle, these
zero-modes are quantized to be integers \cite{Witten}.
As a consequence, the partition function becomes
\begin{eqnarray}
Z &=& \displaystyle{ \sum_{m \in {\bf Z}} } \exp{ i \int d^2 \sigma 
\left[ -\sqrt{1 + m^2} \sqrt{- \det G_{ij}} 
+ C_{01} - m b_{01} \right] },
\label{3.6}
\end{eqnarray}
from which we can read off the effective action
\begin{eqnarray}
S = \int d^2 \sigma \left( -\sqrt{1 + m^2} \sqrt{- \det 
G_{ij}} + C_{01} - m b_{01} \right).
\label{3.7}
\end{eqnarray}
Moreover, recalling the relation stemming from (\ref{2.9})
\begin{eqnarray}
\int_{M_2 = \partial M_3} d^2 \sigma 
( C_{01} - m b_{01} ) =  \int_{M_3} 
( \Omega_3 - m H_3 ) = -i \int_{M_3} 
\bar{E} \wedge \hat{E} \wedge ( m {\cal K} + {\cal I} ) E,
\label{3.8}
\end{eqnarray}
and then carrying out an orthogonal transformation
\begin{eqnarray}
U^T (m {\cal K} + {\cal I}) U = - \sqrt{1 + m^2} {\cal K},
\label{3.9}
\end{eqnarray}
with an orthogonal matrix 
$U = \frac{1}{\sqrt{1 + (m - \sqrt{1+m^2})^2}}[(m - \sqrt{1+m^2})1
- {\cal E}]$, one finally obtains the action
\begin{eqnarray}
S = -\sqrt{1 + m^2} \left( \int_{M_2} d^2 \sigma   \sqrt{- \det 
G_{ij}}  -i \int_{M_3} \bar{E} \wedge \hat{E} \wedge {\cal K} E 
\right).
\label{3.10}
\end{eqnarray}
This is nothing but type IIB Green-Schwarz superstring action
with the modified tension $\sqrt{1 + m^2}$ in a type IIB
supergravity background \cite{Grisaru}. (Note that in \cite{Grisaru}
a complex formalism is used while in this article a real
formalism is used, but by comparison of the two formalisms
we can convince ourselves that the action (\ref{3.10}) corresponds 
to the Green-Schwarz action in the complex formalism.) 

It is worthwhile to notice that the result obtained above agrees with
the tension formula for the $SL(2,Z)$ S-duality spectrum
of strings in the flat background \cite{Schwarz} provided 
that we identify the
integer value $\pi^1 = m$ as corresponding to the $(m,1)$
string. To show more clearly that the tension
at hand is the $SL(2,Z)$ covariant tension, it
would be more convenient to start with the following classical action
\begin{eqnarray}
S = -  n \int_{M_2} d^2 \sigma 
\left[ e^{-\phi} \bigl( \sqrt{- \det ( G_{ij} + {\cal F}_{ij} )}
- C_2 \bigr)
+ \frac{1}{2} \epsilon^{ij} \chi F_{ij} \right],
\label{3.11}
\end{eqnarray}
where $n$ is an integer, and we have introduced the constant 
dilaton $\phi$ and the constant axion $\chi$.
Then following the same path
of derivation as above, we can obtain the manifestly $SL(2,Z)$
covariant tension
\begin{eqnarray}
T = \sqrt{ (m + n \chi)^2 +  n^2 e^{-2\phi}}.
\label{3.12}
\end{eqnarray}

Here we would like to comment two important points. One point is 
that we have shown that there exists $SL(2,Z)$ S-duality in type
IIB superstring theory even in a general type IIB supergravity 
background without reference to any approximation. Thus this relation
is quantum-mechanically exact.

The other point is the problem of whether one can interpret the
orthogonal transformation (\ref{3.9}) as the $SO(2)$ rotation of 
the $N=2$ spinor coordinates. 
In our previous paper \cite{Oda1, Oda2} this problem
was emphasized too much, but on reflection it turns out that this 
problem is rather trivial by the following reason. 
Notice that the torsion constraint in (\ref{2.6}) is obviously 
invariant 
under this rotation. Moreover,
since we require that the original super D-string action and the
fundamental Green-Schwarz action reduce to the well-known forms
of the corresponding flat space actions in the flat space limit,
$E^{I\alpha}$ with the $SO(2)$ index $I$ and $E^a$ must take the
following forms at the lowest order expansion with respect to
the spinor coordinates $\theta$
\begin{eqnarray}
E^I_i &=& \partial_i \theta^I + \ldots, \nn\\
E^a_i &=&  \partial_i X^a - i \bar{\theta}^I \gamma^a \partial_i 
\theta^I + \ldots, 
\label{3.13}
\end{eqnarray}
where the dots indicate the higher order terms reflecting the 
curved nature of the background metric. 
These facts mean that $E^I$ transforms as the adjoint representation 
of the $SO(2)$ group, on the other hand, $E^a$ must be invariant
under an $SO(2)$ rotation. Accordingly, we
can understand that the orthogonal transformation (\ref{3.9})
is indeed performed by an $SO(2)$ rotation of the $N=2$ spinor 
coordinates. 
In this way, we have succeeded in deriving the $SL(2,Z)$
S duality of type IIB superstring theory in type IIB on-shell
supergravity background at least within the 
present context. 

To close this section, let us discuss the case of a general
non-constant dilaton superfield. We will see below that even
if a formulation becomes a little complicated compared with
the case of the vanishing (or constant) dilaton, the essential
point remains unchanged. In this case, the relevant constraints
are given by \cite{Cederwall2}
\begin{eqnarray}
T_{\alpha\beta}^c &=& 2i \gamma_{\alpha\beta}^c, \nn\\
H_{a \alpha\beta} &=& - 2i e^{\frac{1}{2} \phi} 
({\cal K} \gamma_a)_{\alpha\beta}, \nn\\
H_{a b \alpha} &=&  e^{\frac{1}{2} \phi} 
(\gamma_{ab} {\cal K} \Lambda)_{\alpha}, \nn\\
R_{(1) \alpha} &=& 2 e^{-\phi} 
({\cal E} \Lambda)_{\alpha}, \nn\\
R_{(3) a \alpha\beta} &=& 2i e^{-\frac{1}{2} \phi} 
({\cal I} \gamma_a)_{\alpha\beta}, \nn\\
R_{(3) a b \alpha} &=& - e^{-\frac{1}{2} \phi} 
({\cal I} \gamma_{ab} \Lambda)_{\alpha}, 
\label{3.14}
\end{eqnarray}
where $\Lambda_{\alpha} = \frac{1}{2} \partial_{\alpha} \phi$.
These constraints give rise to 
the NS-NS 3-form superfield $H_{3}$ and the 3-form
Wess-Zumino term $\Omega_3$ whose concrete expressions are
of form
\begin{eqnarray}
H_{3} &=& d b_2 \nn\\
&=& i e^{\frac{1}{2} \phi} \bar{E} \wedge \hat{E} \wedge {\cal K} E
+ \frac{1}{2} e^{\frac{1}{2} \phi} \bar{E} \wedge \gamma_{ab} {\cal K}
\Lambda E^b \wedge E^a, \nn\\
\Omega_3 &=& d (C_2 + C_0 {\cal F}) \nn\\
&=& - i e^{-\frac{1}{2} \phi} \bar{E} \wedge \hat{E} \wedge {\cal I} E
- \frac{1}{2} e^{-\frac{1}{2} \phi} \bar{E} \wedge \gamma_{ab} {\cal I}
\Lambda E^b \wedge E^a + 2 e^{-\phi} \bar{E} {\cal E} \Lambda \wedge
{\cal F}.
\label{3.15}
\end{eqnarray}
Here notice that the constraints (\ref{3.14}) and the 3-form superfields
(\ref{3.15}) exactly reduce to the previous expressions (\ref{2.6})
and (\ref{2.8}), respectively when the dilaton superfield $\phi$
is vanishing.

If we start with the classical action \cite{Cederwall2}
\begin{eqnarray}
S &=& S_{DBI} + S_{WZ}, \nn\\
S_{DBI} &=& - \int_{M_2} d^2 \sigma e^{-\frac{1}{2} \phi}
\sqrt{- \det ( G_{ij} + e^{-\frac{1}{2} \phi}{\cal F}_{ij} )}, \nn\\
S_{WZ} &=& \int_{M_2 = \partial M_3} (C_2 + C_0 {\cal F})
= \int_{M_3} \Omega_3,
\label{3.16}
\end{eqnarray}
then the same argument as in the previous section leads to
a dual action
\begin{eqnarray}
S &=& - \int_{M_2} d^2 \sigma e^{-\frac{1}{2} \phi} 
\sqrt{1 + (\lambda + C_0)^2 e^{2 \phi}}   \sqrt{- \det G_{ij}} \nn\\ 
& & {} + \int_{M_3} e^{-\frac{1}{2} \phi} 
\sqrt{1 + (\lambda + C_0)^2 e^{2 \phi}}
\left( i \bar{E} \wedge \hat{E} \wedge {\cal K} E 
+ \frac{1}{2} \bar{E} \gamma_{ab} {\cal K} \Lambda \wedge
E^b \wedge E^a \right),
\label{3.17}
\end{eqnarray}
where $\lambda$ is a constant scalar superfield.
In this way we can also show the $SL(2,Z)$ S-duality in the general
non-constant dilaton superfield.

\section{Discussions}

In this paper, we have studied the property of the $SL(2,Z)$ 
duality of a supersymmetric and $\kappa$-symmetric D-string 
action in a general type IIB on-shell supergravity background. 
It has clearly shown that the $SL(2,Z)$ duality of type
IIB superstring theory found originally in a flat space-time
background holds true even in on-shell supergravity background,
which was anticipated previously but has not proved so far.
This fact is quite illuminating from the viewpoint that 
the $SL(2,Z)$ duality is expected to be an exact symmetry 
of the underlying fundamental theory \cite{Hull} as mentioned
in the introduction. Maybe, one of the most challenging studies
in future would be to promote this global 
discrete symmetry to the local
gauge symmetry, from which we could understand the relation
between the strong coupling phase and the weak coupling phase,
and the Kaluza-Klein compactification e.t.c.

Moreover, we have spelled out the problem of the $SO(2)$
rotation of the $N=2$ spinor coordinates. Our proof utilizes
only an invariance of the constraints and the boundary condition
in the flat background limit so that it can be applied to
the other situations in a straightforward way. 

In a coming longer paper \cite{Oda3}, we will prove 
various duality relations in detail where D2 vs. M2
duality, the self-duality of D3-brane and D4 vs. M5
duality are discussed in a general on-shell supergravity
background. As a final goal, it would be wonderful
to prove the existence of duality transformations in
the background independent matrix models in future
\cite{Oda5}.

\vs 1
\begin{flushleft}
{\bf Acknowledgement}
\end{flushleft}
We are grateful to T. Kimura and M. Tonin for valuable
disussions and M. Cederwall for useful information on
a related work \cite{Cederwall3}. This work was supported in part 
by Grant-Aid for Scientific Research from Ministry of Education, 
Science and Culture No.09740212.

\vs 1


\begin{thebibliography}{99}
\bibitem{Schwarz}
J.H. Schwarz, {\PL{\bf B360} (1995) 13};
ERRATUM ibid. {{\bf B364} (1995) 252.}
\bibitem{GS}
M.B. Green and J.H. Schwarz, {\PL{\bf B136} (1984) 367.}
\bibitem{Schwarz2}
J.H. Schwarz, {\NP{\bf B226} (1983) 269.}
\bibitem{Howe}
P.S. Howe and P.C. West, {\NP{B238} (1984) 181.}
\bibitem{Hull}
C.M. Hull and P.K. Townsend, {\NP{B348} (1995) 109};
E. Witten, {\NP{B443} (1995) 85}.   
\bibitem{Cederwall1}
M. Cederwall, A. von Gussich, B.E.W. Nilsson, P. Sundell
and A. Westerberg, {\NP{\bf B490} (1997) 179, 
hep-th/9611159.}
\bibitem{Bergshoeff}
E. Bergshoeff and P.K. Townsend, {\NP{\bf B490} (1997) 145, 
hep-th/9611173.}
\bibitem{Aganagic1}
J. Aganagic, J. Park, C. Popescu, and J.H. Schwarz, {\NP{\bf B496}
(1997) 215, hep-th/9702133.}
\bibitem{Cederwall2}
M. Cederwall, A. von Gussich, B.E.W. Nilsson, 
and A. Westerberg, {\NP{\bf B490} (1997) 163, 
hep-th/9610148.}
\bibitem{Oda1}
I. Oda, {\PL{\bf B444} (1998) 127, hep-th/9809076.}
\bibitem{Oda2}
I. Oda, {\JHEP{\bf 10} (1998) 015, hep-th/9810024.}
\bibitem{Kimura}
T. Kimura, {hep-th/9810136.}
\bibitem{Oda3}
I. Oda, {\it{Duality of Super D-brane Actions in Type II
Supergravity Background}}, to appear.
\bibitem{de Alwis}
S.P. de Alwis and K. Sato, {\PR{\bf D53} (1996) 7187,
hep-th/9601167.}
\bibitem{Oda4}
I. Oda, {\PL{\bf B430} (1998) 242, hep-th/9802152.}
\bibitem{Witten}
E. Witten, {\NP{\bf B460} (1996) 335, hep-th/9510135.}
\bibitem{Grisaru}
M. Grisaru, P. Howe, L. Mezincescu, B. Nilsson
and P.K. Townsend, {\PL{\bf B162} (1985) 116.}
\bibitem{Oda5}
I. Oda, {\MPL{\bf A13} (1998) 203, hep-th/9709005};
{\NP{\bf B516} (1998) 160, hep-th/9710030}; 
{\PL{\bf B427} (1998) 267, hep-th/9801051}; 
{hep-th/9801085}; {hep-th/9806096, to appear
in {\it Journal of Chaos, Solitons and Fractals}.}
\bibitem{Cederwall3}
M. Cederwall and P.K. Townsend, {\JHEP{\bf 09} (1997) 03, 
hep-th/9709002.}
\end{thebibliography}
\end{document}